\documentclass[useAMS,usenatbib]{mn2e}
\def\aap{AA}

\def\apjl{ApJL}

\def\procspie{Proc. SPIE}
\def\mnras{MNRAS}
\def\apj{ApJ}

\def\pasp{PASP}

\def\Jtwsx{{\rm SDSS J}1206$+$4332}
\def\Reff{R_{\mathrm{e}}}
\def\RE{R_{\mathrm{E}}}
\def\kext{\kappa_{e}}
\def\tdist{D_{\Delta t}}
\newcommand\ion[2]{#1$\;${\scshape{#2}}}%

\usepackage{graphicx}
\usepackage{float}
\usepackage{amssymb}
\usepackage{amsfonts}
\usepackage{amsmath} 
\usepackage{color}

\def\ucsb{Department of Physics, University of California, Santa Barbara, CA 93106, USA}
\def\ucla{Department of Physics and Astronomy, PAB, 430 Portola Plaza, Box 951547, Los Angeles, CA 90095-1547, USA}
\def\ucd{Department of Physics, University of California Davis, 1 Shields Avenue, Davis, CA 95616, USA}
\def\asiaa{Institute of Astronomy and Astrophysics, Academia Sinica, P.O.~Box 23-141, Taipei 10617, Taiwan}
\def\ioa{Institute of Astronomy, Madingley Road, Cambridge CB3 0HA, UK}

\def\aaemail{\tt aagnello@astro.ucla.edu, tt@astro.ucla.edu}

\title[J1206: source, lens and cosmography]
{High resolution imaging and spectroscopy of the gravitational lens SDSSJ1206+4332: a natural coronagraph at $z=1.789$ and a standard ruler at $z=0.745$} 
\author[Agnello et al.]{Adriano Agnello$^{1}$\thanks{\aaemail},
 Alessandro Sonnenfeld$^{1}$,
 Sherry H. Suyu$^{2}$,
 Tommaso Treu$^{1,\dag}$, \and Christopher D. Fassnacht$^{3}$,
 Charlotte Mason$^{4}$, Maru\v{s}a Brada\v{c}$^{3}$, Matthew W. Auger$^{5}$
  \medskip\\
  $^1$\ucla\\
  $^2$\asiaa\\
  $^3$\ucd\\
  $^4$\ucsb\\
  $^5$\ioa\\
  $^\dag$ Packard Fellow.\\
}

\begin{document}

\voffset-.6in

\date{Accepted . Received }

\pagerange{\pageref{firstpage}--\pageref{lastpage}} 

\maketitle

\label{firstpage}

\begin{abstract}
We present spectroscopy and laser guide star adaptive optics (LGSAO)
images of the doubly imaged lensed quasar {\Jtwsx}. We revise the
deflector redshift proposed previously to $z_{d}=0.745,$ and measure
for the first time its velocity dispersion $\sigma=(290\pm30)$
km/s. The LGSAO data show the lensed quasar host galaxy stretching
over the astroid caustic thus forming an extra pair of merging images,
which was previously thought to be an unrelated galaxy in seeing
limited data.  Owing to the peculiar geometry, the lens acts as a
natural coronagraph on the broad-line region of the quasar so that
only narrow [\ion{O}{III}] emission is found in the fold arc. We use
the data to reconstruct the source structure and deflector potential,
including nearby perturbers. We reconstruct the point-spread function
(PSF) from the quasar images themselves, since no additional point
source is present in the field of view. From gravitational lensing and
stellar dynamics, we find the slope of the total mass density profile
to be $\gamma^{\prime}=-\log\rho/\log r =1.93\pm0.09.$ We discuss
the potential of {\Jtwsx}~for measuring time delay distance (and thus
H$_0$ and other cosmological parameters), or as a standard ruler, in
combination with the time delay published by the COSMOGRAIL
collaboration.  We conclude that this system is very promising for
cosmography. However, in order to achieve competitive precision and
accuracy, an independent characterization of the PSF is
needed. Spatially resolved kinematics of the deflector would reduce
the uncertainties further. Both are within the reach of current
observational facilities.
\end{abstract}


\section{Introduction}

In the past two decades, the dream of \citet{Ref64} to use
gravitational time delays as a tool for cosmography has become a
reality. Several observational and modeling breakthroughs now make it
possible to measure time delay distances with 5-6\% precision
and accuracy from a single galaxy scale gravitational lens system
\citep[e.g.][]{suy10a}, and the angular diameter distance to the deflector \citep{par09,jee14}. 
In turn, these absolute distance measurements enable the determination
of the Hubble Constant \citep[e.g.,][]{sch97,tk02,wuc04,jac07,ogu07}, and other
cosmological parameters like flatness and equation of state of dark
energy when combined with other probes \citep{suy13,suy14}. The
necessary observational ingredients include determination of time
delays with few percent precision \citep{bur02,fas02,eul13,tew13}, imaging and astrometry at 10-100 mas resolution
\citep{fas02,koo03,cha10}, stellar velocity dispersion of the 
deflector \citep{tk02}, and imaging and spectroscopy to characterize
the line of sight and environment \citep{kz04,fas11,gre13,col13}. 

The application of lensed quasars for cosmography has been limited by
the small number of systems for which the full complement of data is
available. We present here results from on ongoing program, aimed at
increasing the statistical power of the method, by obtainining all the
required imaging and spectroscopic data for gravitational lensed
quasars with measured time delays.

In this paper, we focus on a very unusual and remarkable system, the
doubly lensed quasar {\Jtwsx}. It was discovered by \citet{ogu05}, who
suggested a redshift of $z=0.748$ for the lens galaxy based on
\ion{Mg}{II} absorption on the quasar spectra, and has been monitored
by \citet{eul13} to obtain a time-delay $\Delta t=111\pm3$ days
between the two quasar images. In those observations, an additional
galaxy `G3' was detected along the line of sight NE of the system,
which would complicate the lens modelling. Our laser guide star
adaptive optics \citep[LGSAO,][]{wiz06} data show that `G3' is in fact
two merging images of lensed quasar host galaxy itself. In practice,
while the quasar is just outside the astroid caustic and thus only
doubly imaged, the host galaxy cross the caustics and produces four
images in a fold configuration. This configuration makes this system
very interesting for cosmography, since it is relatively easy to
monitor as a double, and yet it has a larger number of imaging
constraints available for reconstructing the lensing potential than a
typical pure double.

We complement our LGSAO data with long-slit spectroscopy of the
deflector, whence we can revisit the redshift and provide the first
measurement of the velocity dispersion. Spectroscopy and
high-resolution imaging enable the reconstruction of source structure
and inference of the gravitational lens potential. Together with the
published time-delay, they enable us to assess the role of {\Jtwsx}~
as a probe of the Hubble constant, or alternatively as a standard
ruler.
 
This paper is structured as follows. In Section \ref{sect:AOimg} we
describe the imaging data, complemented with the spectroscopic
follow-up in Section \ref{sect:spec}. In Section \ref{sect:lensmod},
we illustrate the reconstruction of deflector, source and PSF. We add
information form stellar kinematics in Section \ref{sect:dyn},
obtaining an independent characterization of the deflector total mass
density profile. We then discuss the value of {\Jtwsx}~ for
cosmography, in terms of inference on $H_0$ and deviations from the
flat $\Lambda$CDM paradigm. We conclude in Section~\ref{sect:final}.

\begin{figure}
       \centering
       \includegraphics[width=\linewidth]{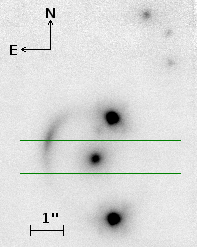}

\caption{\small{
Imaging of {\Jtwsx}~ with LGSAO-NIRC2; North is up and East is left,
the bar marks $1''$.  The two quasar images, the deflector galaxy
\citep[`G1' in][]{ogu05,eul13} and the lensed host galaxy are all
clearly visible. The lensed host traces a typical \textit{fold}
configuration, previously dubbed `G3' and mistaken in the past for an
additional galaxy along the line of sight. Three faint objects are
visible to the NW, at $\approx4.4''$ from the deflector, dubbed `G2'
in previous observations where they were blended into a single, broad and
faint object. The green horizontal lines mark the extent of the DEIMOS
slit used for spectroscopy.}}
\label{fig:coadd}
\end{figure}

\section{Keck Adaptive-Optics Imaging}\label{sect:AOimg}

We imaged {\Jtwsx} in the K$'$ band at $2.2\mu$m with the NIRC2 camera
on Keck II on 2014 January 8th, using the LGSAO system. A total
exposure time of 4860 s was obtained with 27 exposures of 180 s
each. Data were reduced and coadded with pipelines developed by
\citet{aug08}. High resolution is obtained using the pixel scale
of $9.94\times10^{-3}$ arcseconds of the Narrow Camera of NIRC2.

The coadded data are shown in Figure~\ref{fig:coadd}, rebinned in
three-by-three pixels to highlight low surface-brightness
features. Here and through the rest of this paper, North is up and
East is left.  With this resolution, the two quasar images are clearly
visible and distinguished from the deflector galaxy in the middle. The
(lensed) host galaxy is visible around the quasar images and also in
the \textit{fold} arc at NE, which was previously mistaken for a
distinct blue galaxy \citep[dubbed `G3';][]{ogu05,eul13}.
\begin{figure}
       \centering
        \includegraphics[height=0.4\linewidth]{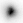}
        \includegraphics[height=0.4\linewidth]{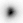}\\
        \includegraphics[height=0.4\linewidth]{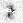}
        \includegraphics[height=0.4\linewidth]{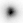}
\caption{\small{
 Cutouts around the two point-sources (top line), their scaled difference (bottom left) and first guess fo the PSF (bottom right) as decribed in Section~\ref{sect:PSFrec}.
}}
\label{fig:psfguess}
\end{figure}

We fit the deflector light with a flattened S{\'e}rsic profile, within a circular patch avoiding contamination from the Einstein ring light. To this aim, the model profile must be convolved with the point-spread function (PSF). Given the small field of view, there is no independent point source at our disposal for this purpose, so we must infer the PSF directly from the system itself. This is described in the subsection~\ref{sect:PSFrec}. The resulting parameters of the deflector light are given in Table~\ref{tab:galpars}.
 We will use the best-fitting light-profile in the following Sections to model the lensing and dynamics.

\begin{table} 
\centering
\begin{tabular}{|c|c|c|c|c|c|}
\hline
$b/a$ & p.a. (N of W) & $n$ & $R_{\rm{eff}}$ &  $\Delta\rm{ra}$  & $\Delta\rm{dec}$   \\
\hline
$0.89$ & $51.24$ deg & 2.66 & $0.60^{\prime\prime}$ & $0.252^{\prime\prime}$ & $2.583^{\prime\prime}$\\
\hline
\end{tabular}
\caption{
Parameters of the best-fitting deflector light profile. From left to right: axis ratio, position angle (counterclockwise, N of W); S{\'e}rsic index; effective radius; relative astrometry from the southern-most quasar image, peak to peak, with r.a. (resp. dec.) increasing to the East (resp. North).}
\label{tab:galpars}
\end{table}
\subsection{PSF estimation}\label{sect:PSFrec}

For the deflector light subtraction, a first guess of the PSF is sufficient. We obtain it by
 isolating small cutouts centred on the two quasar images, subtracting to each of them the average flux around the border, normalizing their fluxes to unity, then coadding the two cutouts.

To recover a PSF that is common to both images and avoid propagating noise, we use a crude version of regularization. For each pixel of the target PSF coadd we consider the two corresponding pixel-values $p_{1},p_{2}$ in the point-source cutouts: when these are within 0.1 of the average $av_{p}=(p_{1}+p_{2})/2,$ we register $av_{p}$ in the PSF coadd, otherwise we register the minimum $\mathrm{min}(p_{1},p_{2}).$ The result is shown in Figure~\ref{fig:psfguess}.

The reconstruction relies on the fact that the lensed host is stretched tangentially by gravitational lensing, so that
 in first approximation it contributes a uniform background flux to the point-source, which should dominate over the PSF light at large enough distances from the PSF core. A variant of this procedure will be combined iteratively with the lens model in Section~\ref{sect:lensmod}.

\section{Keck Spectroscopy}\label{sect:spec}
\begin{figure}
 \centering
 \includegraphics[width=0.5\textwidth]{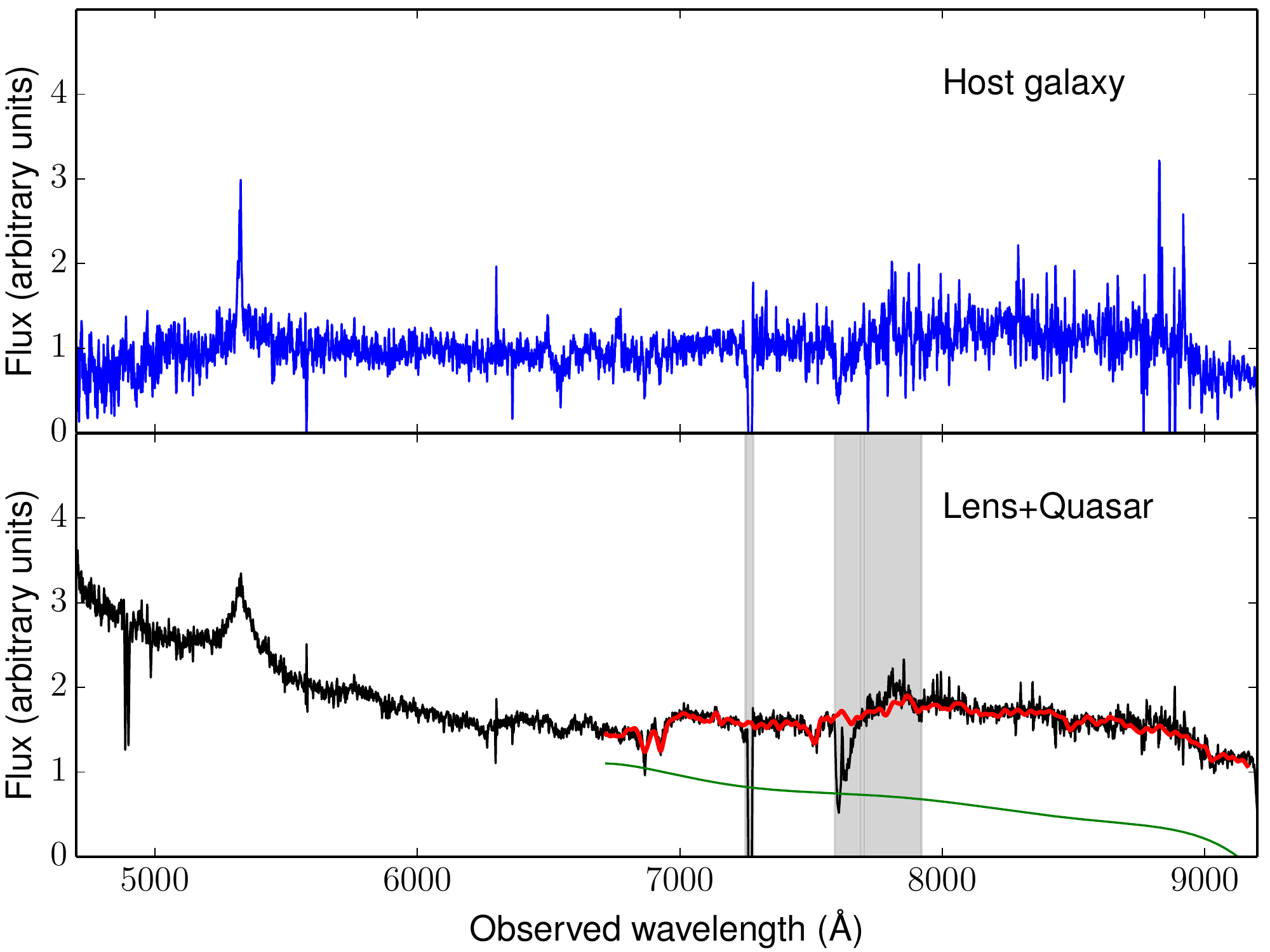}
\caption{\small{
DEIMOS slit spectra. {\em Top:} Extracted 1D spectrum of the quasar host galaxy.
 {\em Bottom:} Extracted 1D spectrum of the quasar (northern point source) and the deflector galaxy.
The red line is the best-fit spectrum obtained from the velocity dispersion fitting.
The green line is the best-fit polynomial used to model continuum emission from the quasar.
The vertical gray bands are regions of the spectrum masked out from the fit.
}}
\label{fig:spec}
\end{figure}

Long-slit spectroscopic measurements of {\Jtwsx}~ were taken with the
instrument DEIMOS \citep{deimos} on Keck 2, on May 16 2015.  The
1''-wide slit was centred on the deflector galaxy and aligned with the
long axis in the E-W direction (see Fig.~\ref{fig:coadd}).\\ We used
the 600ZD grism, covering the wavelength range 4600\AA~-~9200\AA~,
with a spectral resolution of about 160~km~s$^{-1}$ FWHM.  The total
integration time was 1 hour.  We detect a signal from the deflector
galaxy, the quasar, and the quasar host.  The lens galaxy and quasar
host are spatially resolved, while the quasar itself, not covered by
the slit, shows up as a contamination to the lens galaxy spectrum.
The 1D spectra of the lens galaxy and quasar host are plotted in
Figure~\ref{fig:spec}.

The quasar component is visible with its continuum emission in the
blue side of the lens galaxy spectrum and with broad emission lines.
The host galaxy spectrum is very faint. The only clear line detected
is [\ion{C}{III}] 1908\AA, also seen in the quasar component.  However,
while the quasar line is broad (FWHM $\sim6000$~km~s$^{-1}$), the same
line is much narrower in the spectrum of the host.  Thus, we infer
that the quadruply imaged part of the host galaxy does not include the
compact broad line region, but only the significantly more extended
narrow line region (NLR).  In other words, the lens is acting as a
natural ``coronagraph'', blocking out the light of the accretion disk
and the BLR in two of the images.  From a fit to the narrow line
wavelength, we obtain a source redshift $z_{s}=1.789,$ confirming the
one measured by \citet{ogu05}. At lower signal-to-noise ratio,
 there is an absorption feature compatible with FeII at 2344\AA~ at the host redshift $z_{s},$ which is seen also in the quasar spectrum.
 
The lens galaxy spectrum has some prominent stellar absorption
features: Ca K,H at 3934\AA, 3967\AA, the G-band absorption complex
around 4300\AA, and Mgb at 5175\AA, at a redshift $z_d=0.745$.  In
addition to these stellar absorption lines we detect nebular
absorption in the \ion{Mg}{II} doublet 2795\AA~-~2803\AA~ at $z=0.748$.
These lines were also detected by \citet{ogu05}, who used them to
estimate the lens redshift.  There is a rest-frame velocity difference of about
$516$~km~s$^{-1}$ between stellar and \ion{Mg}{II} absorption. We note that this difference in
redshift would introduce a small systematic bias in cosmological
measurements (e.g. 0.15\% in angular diameter distance to the
deflector) based on this system if not corrected.
 
From the data at hand we can also extract a velocity dispersion
measurement.  We do this by fitting stellar templates to the observed
spectrum, using an adaptation of the velocity dispersion fitting code
by \citet{vdm94} 
 described by \citet{suy10a}.  We also fit for continuum emission from the quasar
with a 7-order polynomial.  We use G and F star templates from the
Indo-US library.  The 1D spectrum of the lens is extracted from a
1''$\times$1.9'' rectangular aperture.  We fit the rest-frame
wavelength region between 3800\AA~ and 5250\AA, where the most
important stellar aborption features are found, masking out the region
contaminated by broad MgII emission from the quasar as well as the
deep atmospheric feature around 7600\AA.  The deepest
stellar absorption lines, Ca K,H, fall in correspondance with the
telluric absorption feature at 6900\AA.  We correct for this
atmospheric absorption using a telluric standard star observation as a
reference and keep Ca K,H in the region of the spectrum used for the
fit.  The best-fit spectrum is overploted in Figure~\ref{fig:spec}.
The median signal-to-noise ratio in the region used for the fit is
21\AA$^{-1}$, though if we consider only the signal coming from the
lens galaxy, i.e. once contamination from the quasar continuum
emission is removed, it decreases to 13\AA$^{-1}$ (rest-frame).

We measure a velocity dispersion of $\sigma=290\pm30$~km~s$^{-1}$.
The uncertainty reflects both statistical and systematic errors. We
estimated the latter by repeating the fit over different wavelength
ranges and by changing the order of the polynomial continuum.

\section{Lens Model}\label{sect:lensmod}
\begin{table*} 
\centering
\begin{tabular}{|c|c|c|c|c|c|c|c|c|}
\hline
model  &  $\theta_{\rm{E},l}$  &  $q$  &  $\varphi_{l}$  &  $\gamma_{s}$  &  $\varphi_{s}$  &  $\theta_{\rm{E},p}$  & $\Delta\rm{ra}$ & $\Delta\rm{dec}$ \\
  &  (arcsec)  &    &  [rad]  &    &  [rad]  &  (arcsec)  & (arcsec E) & (arcsec N) \\
\hline
SIE  &  1.61  &  0.807  &  0.65  &  ---  &  ---  &  --- &  ---&  ---  \\
SIS+XS  &  1.582  &  ---  &  ---  &  0.05  &  1.95  &  --- &  --- &  --- \\
SIE+XS  &  1.582$^{\ast}$  &  0.814  &  0.87  &  0.04  &  1.60  &  --- &  --- &  --- \\
SIS+SIS  &  1.571  &  ---  &  ---  &  ---  &  ---  &  0.30 &  -2.506 &  3.485 \\
SIE+SIS  &  1.571$^{\ast}$  &  0.847  &  0.485  &  ---  &  ---  &  0.25 &  -2.198 &  3.852 \\
\hline
\hline
all & 1.575$^{\ast}$ &  0.8$^{\ast}$ &  0.75  &  0.034  &  1.12  &  0.25  & -2.352$^{\ast}$  & 3.669$^{\ast}$ \\
all & 1.575$^{\ast}$ &  0.86$^{\ast}$ & 0.82 & 0.025  &  1.56  &  0.18  & -2.352$^{\ast}$  & 3.669$^{\ast}$ \\
all & 1.575$^{\ast}$ &  0.9$^{\ast}$ & 0.91 & 0.032 &  1.89 & 0.17  & -2.352$^{\ast}$  & 3.669$^{\ast}$ \\
all & 1.575$^{\ast}$ &  0.99$^{\ast}$ & 1.10  &  0.05  &  2.2 & 0.16  & -2.352$^{\ast}$  & 3.669$^{\ast}$ \\
\hline
\end{tabular}
\caption{
Model parameters obtained from conjugate-point analysis, using different models. Parameters marked by an asterisk are kept fixed during the optimization. These are then used to initialize a model with main lens, external perturber G2 and external shear, which is required to fit the quasar image positions and arc surface-brightness profile. The Einstein radius of the main lens is kept fixed to its SIS value when generalizing to SIE. The second part of the table shows results for a model with SIE, external shear and perturber, with some paramters fixed.}
\label{tab:init}
\end{table*}

We use the modelling code GLEE \citep{suy10,suy12} to infer the lensing potential and reconstruct the extended source structure.
 As constraints, we use the arc surface-brightness distribution and the point source positions.
 The arc is cleaned from possible contamination by the lens light, by subtracting the best-fitting profile found in Section~\ref{sect:AOimg}.
 Due to uncertainties in the PSF reconstruction, we artificially increase the pixel uncertainties around the point-source images,
 so that the lens model is constrained mainly by the NE fold and by the two quasar image positions,
 to which we assign a positional uncertainty of 20 mas.
 For computational simplicity, we consider just the pixels in an annular mask around the Einstein radius.
 
Inference on the lens mass profile relies primarily on the differential distortion and radial magnification of the same source patch on different sides of the lens. Then, two conditions must be met: (i) high resolution imaging, with good seeing conditions and small pixel size; (ii) a robust characterization of the PSF. While the former is satisfied by our Keck images, the latter is not directly available and the PSF must be reconstructed from the images.

\begin{figure*}
 \centering
 \includegraphics[width=0.19\textwidth]{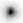}
 \includegraphics[width=0.19\textwidth]{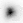}
 \includegraphics[width=0.19\textwidth]{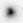}
 \includegraphics[width=0.19\textwidth]{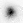}
 \includegraphics[width=0.19\textwidth]{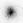}\\
 \includegraphics[height=0.33\textwidth]{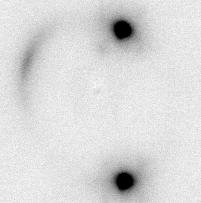}
 \includegraphics[height=0.33\textwidth]{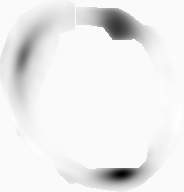}
 \includegraphics[height=0.33\textwidth]{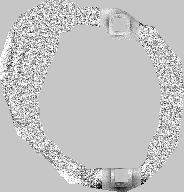}
\caption{\small{ Reconstructed lens model.
 \textit{Top:} iterative corrections of the reconstructed PSF. A 23-by-23 pixel cutout was chosen for the PSF; since the third iteration, the small-scale
  PSF  corrections start to overfit noise in the data.
\textit{Bottom:} deflector-light-subtracted data (left), model of the arc (middle) and residuals over noise (right). Secondary lobes in the PSF are evident in the deflector-light-subtracted image and are masked out when fitting the extended structure. Secondary `wings' in the reconstructed profile around the quasar images (near the mask border) are given by residuals in masking the PSF lobes out and choice of error-map. Their appearance is also degenerate with the point-source fluxes, which can create `holes' in the reconstructed arc.
 The residuals on the fold arc are completely compatible with noise.
}}
\label{fig:LenRes}
\end{figure*}

The full model would then encompass both the lens parameters and the PSF reconstruction. We proceed iteratively, alternating PSF reconstruction and lens model fitting. From a run of the lens model, we get the predicted surface-brightness profile of the lensed host, which we can subtract near the quasar images. Similarly to the procedure in Section~\ref{sect:PSFrec}, we select two cutouts around the (host-subtracted) point-source images and combine their common pixel-values into a new PSF, which is then used in a new run of the lens model. Figure~\ref{fig:LenRes} shows the inferred PSF in subsequent iterations. After three iterations, the correction step starts to overfit noise, while the parameters of the lens model are not changing appreciably.

\subsection{Model Components}

We adopt a power-law profile for the total (i.e. luminous plus dark matter) mass density of the main deflector, with convergence
\begin{equation}
\kappa(x,y)=\frac{3-\gamma^{\prime}}{1+q}\left(\frac{\sqrt{x^{2}+q^{-2}y^{2}}}{\RE}\right)^{1-\gamma^{\prime}}
\end{equation}
along the principal axes of the deflector. With this convention, the Einstein radius $\RE$ is such that in the circular limit ($q=1$) it encloses a mean convergence of 1. In what follows, the model Einstein radii will be in angular units and denoted by $\theta_{E}.$

The main deflector has a three nearby, faint perturbers to the NW, at a distance of approximately 4.4'', as shown in Figure~\ref{fig:coadd}. In previous observations \citep{ogu05,eul13}, they were all blended into a single, broad and faint clump dubbed `G2'. In order to model contributions from the immediate environment, we then include a Spherical Isothermal Sphere (SIS, $\gamma^{\prime}=2,$ $q=1$) centred on G2, with a centre to be adjusted, plus external shear. The role of shear is to correct for: flattening and/or imperfect centering of the group G2; lesser contributions from other galaxies in the vicinity of the lens; mismatch between the adopted and real flattening of the main deflector.

The model then has the following parameters: the power-law exponent $\gamma^{\prime}$ of the density profile of the deflector, its axis ratio $q$ and position angle $\varphi_{l}$ (N of W); the shear amplitude $\gamma_{s}$ and position angle $\varphi_{s}$ (N of W); the nearest perturber (G2) Einstein radius $\theta_{E,p}$ and its position; the source-position, constrained by the quasar image positions within a 20 mas range, and the point-source fluxes $f_{1},f_{2}$.
 The point-source fluxes are computed but not used for the lens model, as they can be affected by source variability, microlensing, and milli-lensing by possible satellites of the main deflector.
 One final parameter would the Einstein radius $\theta_{E,l}$ of the main deflector, which we keep fixed because it can be determined with 1\% precision as shown below.

\subsection{Initialization and Choice of Parameters}
Initialization is performed via conjugate-point analysis, i.e. identifying sets of points that should map to the same location in the source-plane and fitting for their predicted positions in the image-plane. Given the low number of constraints, we can examine just combinations of separate models, whose parameters will be cross-checked for consistency. The first initialized model has an isolated Singular Isothermal Ellipsoid (SIE, $\gamma^{\prime}=2,$ $q<1$), with free Einstein radius, axis ratio and position angle. The second model has a SIS plus external shear, the third has a SIS for the main deflector and one for G2. Since the SIS models yield consistent measurements of Einstein radius, we then generalize the main deflector to a SIE, keeping its $\RE$ to the SIS value, and optimize a model with external shear and one with a SIS for the external perturber G2.

 The results are shown in Table~\ref{tab:init}. The position of G2 is consistent within $\approx0.4''.$ Its inferred inferred values of $\theta_{E}$ are consistent within 0.05''. The difference in $\theta_{\rm{E,l}}$ between a model with external shear and a model with external SIS perturber is $\approx0.01''.$
 
The interplay between different parameters is explored by different choices of flattening of the main deflector, keeping the perturber position fixed to the average value between the SIS+SIS and SIE+SIS cases. Results from this experiment are shown in the second half of Table~\ref{tab:init}. When $q=0.9$ as derived from the photometry of the deflector galaxy in Sect.~\ref{sect:AOimg}, the inferred deflector p.a. from lensing is exactly the same as the one found from the photometry. This finding suggests that $q=0.9$ and $\varphi_{l}=0.91$ can be kept fixed in the lens model. Similarly, we keep the perturber position fixed, since possible corrections will be given by the shear term. 

 Unfortunately, with the data at our disposal, a complete lens model on the extended source
 is unstable and some of the parameters converge to pathological minima. For this reason, we are currently
 forced to keep some of the parameters fixed to their fiducial values.
\begin{figure}
 \centering
 \includegraphics[width=0.475\textwidth]{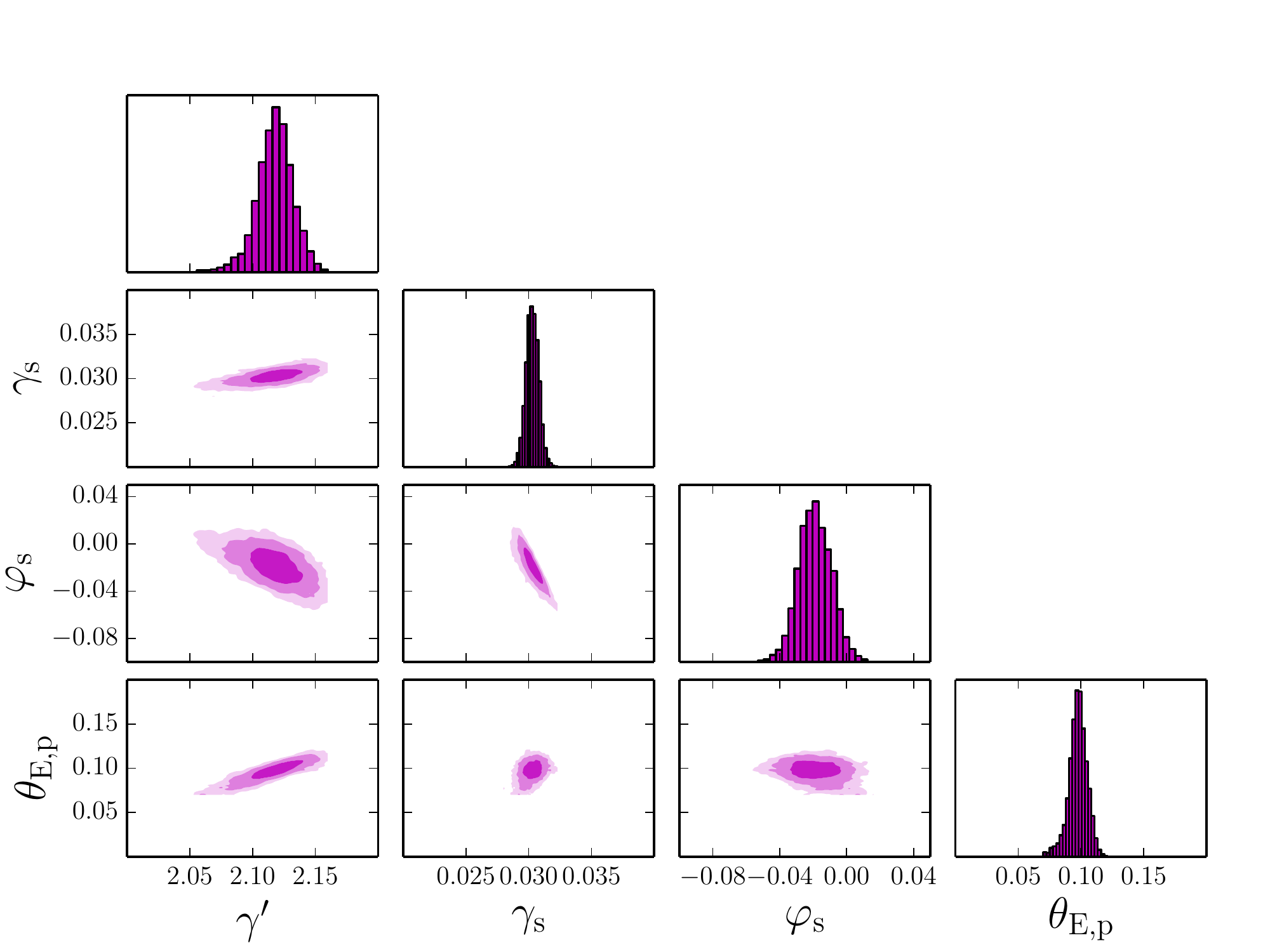}
\caption{\small{ Reconstructed lens model.
 Inference on the four main parameters of the lens model. We show the marginalized posterior distribution of the power-law slope $\gamma^{\prime},$ external shear amplitude $\gamma_{s}$ and angle $\varphi_{s},$
 and the perturber (G2) Einstein radius $\theta_{\rm{E},p}$ in arcseconds. Due to the very small pixel-size and lack of independent PSF, the model uncertainties are artificially small.
}}
\label{fig:corner}
\end{figure}
\subsection{Lensing Results}
Results from the model are shown in Figure~\ref{fig:corner}, showing the inferred posterior distribution on the four main parameters.
Their values are in general agreement with what is found at the initialization stage. The Einstein radius $\theta_{\rm{E},p}$ of G2 is small but non negligible. Other perturbations to the lensing potential are summed up in a small shear, which corroborates our working hypothesis. The density slope $\gamma^{\prime}=2.04$ is in line with the general description of early-type galaxy lenses \citep{koo09},
 but steeper than the median value $\approx1.95$ of lens galaxies at this redshift \citep{son13}.

The statistical uncertainties are driven by the small pixel size compared to the Einstein radius, and are therefore dominated by systematics in the choice of lens model, PSF correction scheme, and estimation of errors near the point-source images. Results on other systems with HST data, with a robustly characterized PSF and deeper observations on the Einstein ring, typically yield uncertainties around $2-3\%$ on $\gamma^{\prime}$ \citep{dye05,suy13}. In what follows, we will adopt an estimated $5\%$ for the uncertainties on $\gamma^{\prime}.$

Galaxies in the vicinity of the main deflector contribute to the deflections of light-rays. As a first approximation, their role can be subsumed in a uniform sheet of convergence $\kext$ around the main deflector. However, this can be reabsorbed by (unobservable) shifts and rescalings in the source-plane, so that $\kext$ cannot be determined from lensing alone. This \textit{mass-sheet degeneracy} \citep[MSD,][]{fal85} is
 a simple case of source-position transformation degeneracies \citep{sch13}, with particular importance for time-delay cosmography.
Bounds on $\kext$ can be put by studying the environment of the lens \citep{fas11,gre13}. \citet{ogu05} have shown that the environment of {\Jtwsx}~ is comparable to a random field of view, considering galaxies within 1 arcmin from the lens down to an $I-$band magnitude of 24. Then, given the current data quality, we can neglect $\kext$ for our purposes. We will discuss this choice further in the following Sections.

\section{Dynamics}\label{sect:dyn}
\begin{figure}
 \centering
 \includegraphics[width=0.45\textwidth]{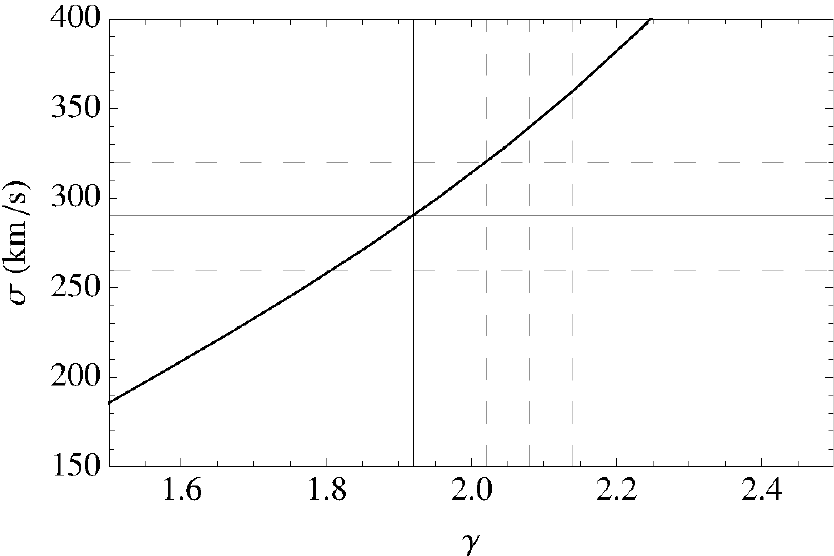}
\caption{\small{ 
Combined inference from lensing and dynamics. The model velocity dispersion is plotted against total density exponent $\gamma^{\prime}=-\mathrm{d}\log\rho/\mathrm{d}\log r,$ corresponding to the measured Einstein radius $\theta_{\rm{E},l}=1.575''.$
 We also show the measured velocity dispersion with its uncertainty (horizontal lines) and the density exponent from the lens model,
 with a conservative uncertainty of $5\%$ (vertical lines).
}}
\label{fig:gammasigma}
\end{figure}
Information on the kinematics of stars in the deflector yields and independent mass probe, thus breaking residual degeneracies in the lens model \citep{tk02,tk04}. This is especially true for the power-law profiles used here, for which simple relations hold between the velocity dispersion and power-law exponent \citep{koo06}. From gravitational lensing, a normalization condition holds for the projected mass of the lens within the Einstein radius:
\begin{equation}
M_{p}(R_{E})=(1-\kappa_{e})\pi R_{E}^{2}\Sigma_{cr}\ =\ \frac{c^{2}D_{s}}{4GD_{l}D_{ls}}(1-\kappa_{e})R_{E}^{2}\ .
\end{equation}
In terms of the mass $M(r),$ surface-brightness $\Sigma(R)=\Sigma_{0}\tilde{\Sigma}(R/\Reff)$ and 3D luminosity $\nu(r)=\tilde{\nu}(r/\Reff)\Sigma_{0}/\Reff$ profiles, the line-of-sight velocity dispersion $\sigma_{p}(R)$ satisfies
 \begin{equation}
 \Sigma(R)\sigma_{p}^{2}(R)\ =\ 2G\int_{R}^{\infty}\frac{\nu(r)M(r)}{r^{2}}(\sqrt{r^{2}-R^{2}}+k_{\beta}(r,R))\mathrm{d}r
 \end{equation}
 where the kernel $k_{\beta}$ depends on the orbital anisotropy profile $\beta(r)$ \citep{mam05,agn14}. In the power-law case,
  with $\rho(r)\propto r^{-\gamma^{\prime}},$ the lensing mass normalization yields
  \begin{eqnarray}
  \label{eq:lensdyn}
 \tilde{\Sigma}(R/\Reff)\sigma_{p}^{2}(R)\ =\ (1-\kappa_{e})c^{2}\frac{D_{s}R_{E}}{D_{ls}D_{l}}\times\frac{(3-\gamma^{\prime})\Gamma(\gamma^{\prime}/2)}{\sqrt{\pi}\Gamma((\gamma^{\prime}-1)/2)}\times\\
 \nonumber \frac{\Reff^{2-\gamma^{\prime}}}{R_{E}^{2-\gamma^{\prime}}}
 \int_{R/\Reff}^{\infty}\tilde{\nu}(x)x^{-\gamma^{\prime}}\left(\sqrt{x^{2}-(R/\Reff)^{2}}+\tilde{k}_{\beta}(x,R/\Reff)\right)\mathrm{d}x
 \end{eqnarray}
 The model velocity dispersion can be averaged over the slit or aperture to be directly compared with its measured value, either via numerical integration or with the more compact formalism of \citet{agn13}, which is adequate here because most of the deflector light is enclosed in the DEIMOS slit. Then, given $\theta_{E}=R_{E}/D_{l}$ and the distance-ratio $D_{s}/D_{ls},$ from the measured kinematics one can derive the exponent $\gamma^{\prime}$ and \textit{vice versa}. In Figure~\ref{fig:gammasigma} we show the predicted velocity dispersion as a function of $\gamma^{\prime},$ together with the measured value and the density exponent inferred from the lens model.

Figure~\ref{fig:gammasigma} shows the predicted velocity dispersion $\sigma$ as a function of $\gamma^{\prime}.$ The vertical lines delimit the 68\% confidence interval in $\gamma^{\prime}$ predicted by the lens model, while the horizontal lines show the measured
 $\sigma$ with its 68\% confidence interval. The data are not sufficient to discriminate between models with different anisotropy, since we just have an aperture-averaged value (with broad PSF) instead of a radial profile at our disposal.
 
The slope obtained this way, $\gamma^{\prime}=1.93\pm0.09,$ is smaller than the value inferred from lensing alone. Still, when a conservative uncertainty of $5\%$ on the lensing $\gamma^{\prime}$ is adopted, the two results are compatible within $68\%$ confidence level.

\begin{figure*}
 \centering
 \includegraphics[width=0.95\textwidth]{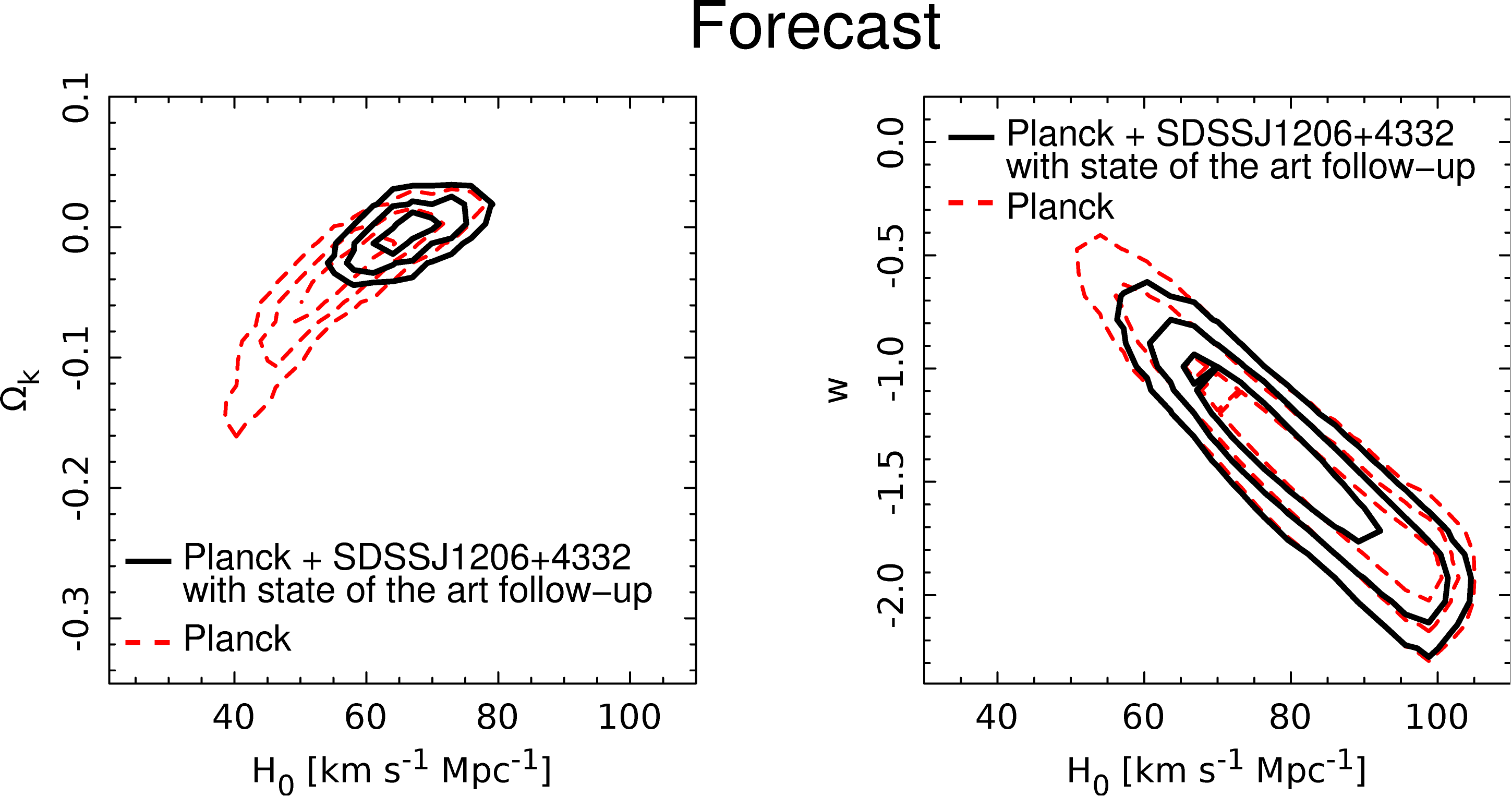}
\caption{\small{
{\bf Predicted} inference on cosmological parameters, in presence of robust model parameters (with well controlled systematics) from more reliable data than the ones available here. Dashed red lines are the Planck prior on cosmological parameters \citep{pla14}, the black lines are the marginalized posterior using {\Jtwsx} as a cosmographic probe, {\it assuming state of the art follow-up data are available}. Degeneracies between $H_{0}$ and other cosmological parameters are appreciably reduced by time-delay lenses, which probe primarily $H_0.$ \textit{Left:} Open $\Lambda$CDM cosmology (free curvature parameter $\Omega_{k}$).
 \textit{Right:} Flat CDM ($\Omega_{k}=0$) with free equation of state parameter $w.$
}}
\label{fig:cosmo}
\end{figure*}
\section{Prospects for Cosmography}
The time-variability of the source quasar provides an additional observable, which is the time-delay between the two quasar images.
 The arrival time is given by
\begin{equation}
t\ =\ (1+z_{d})\frac{D_{s}D_{d}}{cD_{ds}}\left[(\mathbf{x}_{I}-\mathbf{x}_{S})^{2}-\psi(\mathbf{x}_{I})\right]
\end{equation}
in terms of the projected lens potential $\psi,$ deflection $\mathbf{x}_{I}-\mathbf{x}_{S}$ and cosmological distances.

From the lens model, the predicted time-delay can be compared with its measured value to infer the time-delay distance
$D_{\Delta t}=(1+z_{d})D_{l}D_{s}/D_{ds}\propto H_{0}^{-1},$ or equivalently the cosmological parameters, primarily the Hubble constant. This is generally plagued by the MSD, as the source-position transformation that eliminates $\kext$ from the lens model produces
 a biased measurement $H_{0}/(1-\kext)$ of the Hubble constant,
  with everything else fixed -- cosmological abundances and measured $\Delta t.$
 If the velocity dispersion and time-delay are modelled jointly, the common dependence on $(1-\kappa_{e})D_{s}/D_{ds}$
  cancels out, so that the angular-diameter distance $D_{d}$ to the deflector is inferred directly,
  with possible systematics from $\kappa_{e}$ now erased \citep{jee14}.
  Alternatively, in presence of a good prior on $\kext,$ dynamics can be used to obtain the exponent $\gamma^{\prime}$
  to be used in the time-delay inference, since the model kinematics depend just weakly on cosmological parameters (eq.~\ref{eq:lensdyn}).

Given the data presented here, with small nominal uncertainties on $\gamma^{\prime},$ one would be tempted to exploit them for time-delay cosmography. We caution against that, for the following reasons. First, the signal-to-noise ratio of faint features, such as the fold arc and perturbers in G2, is too low to enable a flexible model with robustly determined parameters. Second, the lack of an independent PSF model leaves degeneracies in the lens model unresolved, degrading the predictive power near the point-source positions. This is particularly true for two-quasar-image systems like {\Jtwsx}; in four-quasar-image systems, there is more information from the four quasars to reconstruct the PSF directly from the lens system (Chen et al., in preparation). The confidence ranges of lensing parameters account just for the statistical uncertainties and are driven mainly by the small pixel-size, whereas systematic uncertainties (e.g. from PSF correction) dominate the error budget.

Bearing the caveats in mind, it is interesting to provide some rough
cosmographic estimates and discuss what can be earned from more
reliable data. For concreteness, we forecast the cosmographic
constraints that we expect to obtain with (1) HST imaging that have
better characterized PSFs, and (2) spatially resolved kinematics of
the deflector galaxy that help break lens-modeling and mass-sheet
degeneracies.  To estimate the precision on the time-delay distance,
we assume uncertainties on the time delays of 3\% (Eulaers et
al. 2013), modeling with HST imaging of 3\% (e.g., Suyu et al. 2013),
and external convergence $\kext$ of 4\% (e.g., Greene et al. 2013,
Collett et al. 2013).  Adding these contributions in quadrature, we
forecast an uncertainty of 5.8\% on $\tdist.$ For reference, we adopt
a fiducial value for $\tdist$ which corresponds to a fiducial flat
$\Lambda$CDM model with $\Omega_{\rm m}=1-\Omega_{\Lambda}=0.3$ and
$H_0=70\,{\rm km\,s^{-1} Mpc^{-1}}.$ Assuming a Gaussian probability
distribution for $\tdist=5789\pm336\, {\rm Mpc}$, we show in
Figure~\ref{fig:cosmo} the cosmographic information from {\Jtwsx}~
assuming Planck priors \citep{pla14} in either open $\Lambda$CDM
cosmology (left panel) and flat $w$CDM cosmology (right panel).  Such
a distance measurement of $\tdist$ for {\Jtwsx} would help break
significantly parameter degeneracies in the CMB data, and provide an
independent test of systematics in cosmographic probes.


\section{Summary}
\label{sect:final}

We have presented laser guide star adaptive optics images and
spectroscopic observations of the gravitational lens system
\Jtwsx.  A precise time delay measurement is available for this 
doubly imaged quasar \citep{eul13}, making it a prime target for
follow-up observations aimed at using it as tool for cosmography.  The
high resolution images allow us to recognize that the system is in a
very unusual and favorable configuration for cosmography, previously
not resolved in seeing based images.  Furthermore, our deep Keck
spectroscopy allows to correct the previous estimate of the redshift
of the main deflector and measure its stellar velocity dispersion
based on stellar absorption features.  We then use the newly available
information to construct a gravitational lens model of the system and
discuss its potential for cosmography. Our main results can be
summarized as follows:

\begin{itemize}
\item The quasar lies just outside of the astroid caustic in the source
plane and is thus doubly imaged. However, the quasar host galaxy
crosses the caustic and produces two additional merging images in a
classic fold configuration -- previously mis-identified as an
unrelated galaxy along the line of sight in seeing limited images. The
presence of two additional images doubles the amount of information
available to constrain the gravitational lens model.
\item The lensing geometry acts as a natural ``coronagraph'', 
blocking out the light coming from the broad line region and the
accretion disk in two of the images. Thus, those two images give a
very clear view of the narrow line region surrounding a $z=1.79$
quasar. This system represents an interesting target for follow-up
studies with an integral field spectrograph.
\item A small group of galaxies is identified near the main deflector and included 
in the mass model. The group of galaxies had previously been
misidentified as a single galaxy in seeing limited images.
\item The redshift of the main deflector is measured to be $z=0.745$ from stellar absorption features, revising the previous estimate pf $z=0.748$ based on an intervening \ion{Mg}{II} absorber.
\item The stellar velocity dispersion of the main deflector is measured to be $290\pm30$ kms$^{-1}$.
\item Modeling the surface brightness distribution of the lensed source requires an accurate model of the PSF. As no stellar image is available in the narrow field of view of NIRC2 we use an iterative procedure to reconstruct the PSF from the lensed quasar images themselves.
\item We construct lens models of the system, including nearby perturbers in addition to the main deflector. The  models reproduce the data for a close-to-isothermal total mass density profile of the main deflector. Some of the lens model parameters suffer from relatively large uncertainties dominated by residual PSF uncertainties.
\item We combine the lensing and dynamical information to estimate the logarithmic slope of the total mass density profile $\gamma'$. We find $\gamma'=1.93\pm0.09$ consistent with typical values found for massive early-type galaxies at these redshifts.
\item We estimate that the current systematic uncertainties of the mass model of this system are too large to provide any interesting cosmographic inference. However, we show that, with the addition of HST images and spatially resolved stellar velocity dispersion, the system would become very useful. As an illustration we combine the simulated lensing likelihood with the Planck prior, and show that the uncertainty on $w$, curvature and H$_0$ is significantly reduced.
\end{itemize}

\section*{Acknowledgments}

AA, AS, TT, and CDF acknowledge support from NSF grants  AST-1312329 and AST-1450141
``Collaborative Research: Accurate cosmology with strong gravitational
lens time delays''. AA, AS, and TT gratefully acknowledge support by
the Packard Foundation through a Packard Research Fellowship to TT.
S.H.S. acknowledges support from the Ministry of Science and
Technology in Taiwan via grant MOST-103-2112-M-001-003-MY3. 

Data presented in this paper were obtained at the W.M. Keck Observatory, which is operated as a scientific partnership among the California Institute of Technology, the University of California, and the National Aeronautics and Space Administration. The Observatory was made possible by the generous financial support of the W. M. Keck Foundation. The authors also recognize and acknowledge the very significant cultural role and reverence that the summit of Mauna Kea has always had within the indigenous Hawaiian community. We are most fortunate to have the opportunity to conduct observations from this mountain.

\label{lastpage}

\begin{thebibliography}{}

\bibitem[Agnello et al.(2013)]{agn13} Agnello, A., Auger, M.~W., \& Evans, N.~W.\ 2013, \mnras, 429, L35 
\bibitem[Agnello et al.(2014)]{agn14} Agnello, A., Evans, N.~W., \& Romanowsky, A.~J.\ 2014, \mnras, 442, 3284 
\bibitem[Auger et al.(2008)]{aug08} Auger, M.~W., Fassnacht, C.~D., Wong, K.~C., et al.\ 2008, \apj, 673, 778
\bibitem[Burud et al.(2002)]{bur02} Burud, I., Hjorth, J., Courbin, F., et al.\ 2002, \aap, 391, 481 
\bibitem[Chantry et al.(2010)]{cha10} Chantry, V., Sluse, D., \& Magain, P.\ 2010, \aap, 522, A95 
 
\bibitem[Collett et al.(2013)]{col13} Collett, T.~E., Marshall, P.~J., Auger, M.~W., et al.\ 2013, \mnras, 432, 679 
\bibitem[Dye \& Warren(2005)]{dye05} Dye, S., \& Warren, S.~J.\ 2005, \apj, 623, 31
\bibitem[Eulaers et al.(2013)]{eul13} Eulaers, E., Tewes, M., Magain, P., et al.\ 2013, \aap, 553, A121 
\bibitem[Faber et al.(2003)]{deimos} Faber, S.~M., Phillips, A.~C., Kibrick, R.~I., et al.\ 2003, \procspie, 4841, 1657 
\bibitem[Falco et al.(1985)]{fal85} Falco, E.~E., Gorenstein, M.~V., \& Shapiro, I.~I.\ 1985, \apjl, 289, L1 
\bibitem[Fassnacht et al.(2011)]{fas11} Fassnacht, C.~D., Koopmans, L.~V.~E., \& Wong, K.~C.\ 2011, \mnras, 410, 2167 
\bibitem[Fassnacht et al.(2002)]{fas02} Fassnacht, C.~D., Xanthopoulos, E., Koopmans, L.~V.~E., \& Rusin, D.\ 2002, \apj, 581, 823 

\bibitem[Greene et al.(2013)]{gre13} Greene, Z.~S., Suyu, S.~H., Treu, T., et al.\ 2013, \apj, 768, 39 
\bibitem[Jackson(2007)]{jac07} Jackson, N.\ 2007, Living Reviews in Relativity, 10, 4 
\bibitem[Jee et al.(2014)]{jee14} Jee, I., Komatsu, E., \& Suyu, S.~H.\ 2014, arXiv:1410.7770 
\bibitem[Keeton \& Zabludoff(2004)]{kz04} Keeton, C.~R., \& Zabludoff, A.~I.\ 2004, \apj, 612, 660 

\bibitem[Koopmans et al.(2002)]{koo02} Koopmans, L.~V.~E., Garrett, M.~A., Blandford, R.~D., et al.\ 2002, \mnras, 334, 39 
\bibitem[Koopmans et al.(2003)]{koo03} Koopmans, L.~V.~E., Treu, T., Fassnacht, C.~D., Blandford, R.~D.,
 \& Surpi, G.\ 2003, \apj, 599, 70 
\bibitem[Koopmans(2006)]{koo06} Koopmans, L.~V.~E.\ 2006, EAS Publications Series, 20, 161 
\bibitem[Koopmans et al.(2009)]{koo09} Koopmans, L.~V.~E., Bolton, A., Treu, T., et al.\ 2009, \apjl, 703, L51 
\bibitem[Mamon \& {\L}okas(2005)]{mam05} Mamon, G.~A., \& {\L}okas, E.~L.\ 2005, \mnras, 363, 705 
\bibitem[McCully et al.(2014)]{mcc14} McCully, C., Keeton, C.~R., Wong, K.~C., \& Zabludoff, A.~I.\ 2014, \mnras, 443, 3631 
\bibitem[Oguri et al.(2005)]{ogu05} Oguri, M., Inada, N., Hennawi, J.~F., et al.\ 2005, \apj, 622, 106 
\bibitem[Oguri(2007)]{ogu07} Oguri, M.\ 2007, \apj, 660, 1
\bibitem[Paraficz \& Hjorth(2009)]{par09} Paraficz, D., \& Hjorth, J.\ 2009, \aap, 507, L49 

\bibitem[P{\'e}rez-R{\`a}fols et al.(2015)]{per15} P{\'e}rez-R{\`a}fols, I., Miralda-Escud{\'e}, J., Lundgren, B., et al.\ 
2015, \mnras, 447, 2784 
\bibitem[Planck Collaboration et al.(2014)]{pla14} Planck Collaboration, Ade, P.~A.~R., Aghanim, N., et al.\ 2014, \aap, 571, A16 
\bibitem[Refsdal(1964)]{Ref64} Refsdal, S.\ 1964, \mnras, 128, 307 
\bibitem[Schechter et al.(1997)]{sch97} Schechter, P.~L., Bailyn, C.~D., Barr, R., et al.\ 1997, \apjl, 475, L85 

\bibitem[Schneider \& Sluse(2013)]{sch13} Schneider, P., \& Sluse, D.\ 2013, \aap, 559, A37
 \bibitem[Sonnenfeld et al.(2013)]{son13} Sonnenfeld, A., Treu, T., Gavazzi, R., et al.\ 2013, \apj, 777, 98 
\bibitem[Suyu et al.(2010)]{suy10a} Suyu, S.~H., Marshall, P.~J., Auger, M.~W., et al.\ 2010, \apj, 711, 201 
\bibitem[Suyu \& Halkola(2010)]{suy10} Suyu, S.~H., \& Halkola, A.\ 2010, \aap, 524, A94
\bibitem[Suyu et al.(2012)]{suy12} Suyu, S.~H., Hensel, S.~W., McKean, J.~P., et al.\ 2012, \apj, 750, 10 
\bibitem[Suyu et al.(2013)]{suy13} Suyu, S.~H., Auger, M.~W., Hilbert, S., et al.\ 2013, \apj, 766, 70 
\bibitem[Suyu et al.(2014)]{suy14} Suyu, S.~H., Treu, T., 
Hilbert, S., et al.\ 2014, \apjl, 788, L35
\bibitem[Tewes et al.(2013)]{tew13} Tewes, M., Courbin, F., Meylan, G., et al.\ 2013, \aap, 556, A22  
\bibitem[Treu \& Koopmans(2002)]{tk02} Treu, T., \& Koopmans, L.~V.~E.\ 2002, \mnras, 337, L6 
\bibitem[Treu \& Koopmans(2004)]{tk04} Treu, T., \& Koopmans, L.~V.~E.\ 2004, \apj, 611, 739 
\bibitem[van der Marel et al.(1994)]{vdm94} van der Marel, R.~P., Rix, H.~W., Carter, D., et al.\ 1994, \mnras, 268, 521 

\bibitem[Wizinowich et al.(2006)]{wiz06} Wizinowich, P.~L., Le Mignant, D., Bouchez, A.~H., et al.\ 2006, \pasp, 118, 297 
\bibitem[Wucknitz et al.(2004)]{wuc04} Wucknitz, O., Biggs, A.~D., \& Browne, I.~W.~A.\ 2004, \mnras, 349, 14 





\end{thebibliography}
\end{document}